\documentclass[%
reprint,
showpacs,
 amsmath,amssymb,
aps,
prb,
]{revtex4-1}

\usepackage{amsfonts}
\usepackage{amsmath, amssymb}
\usepackage{bm} 
\usepackage{graphicx} 
\usepackage{epsfig}
\usepackage{epstopdf}
\usepackage{color}
\usepackage{ulem}
\usepackage{hyperref} 

\begin{document}


\title{Transport and collective radiance in a basic quantum chiral optical model.}

\author{D. F. Kornovan}
\email{newparadigm.dk@gmail.com}
\affiliation{ITMO University, Birzhevaya liniya 14, 199034 St.-Petersburg, Russia}

\author{M.I. Petrov}%
\affiliation{ITMO University, Birzhevaya liniya 14, 199034 St.-Petersburg, Russia}
\affiliation{University of Eastern Finland, Yliopistokatu 7, FI-80101 Joensuu, Finland}

\author{I. V. Iorsh}%
\affiliation{ITMO University, Birzhevaya liniya 14, 199034 St.-Petersburg, Russia}

\date{\today}

\begin{abstract}
In our work, we study the dynamics of a single excitation in an one-dimensional array of two-level systems, which are chirally coupled through a single mode waveguide. The chirality is achieved owing to a strong optical spin-locking effect, which in an ideal case gives perfect unidirectional excitation transport. We obtain a simple analytical solution for a single excitation dynamics in the Markovian limit, which directly shows the tolerance of the system with respect to the fluctuations of emitters position. We also show that the Dicke state, which is well-known to be superradiant, has twice lower emission rate in the case of unidirectional quantum interaction. Our model is supported and verified with the numerical computations of quantum emmiters coupled via surface plasmon modes in a metalic nanowire. The obtained results are based on a very general model and can be applied to any chirally coupled system, that gives a new outlook on quantum transport in chiral nanophotonics. 
\end{abstract}


\maketitle

\section{Introduction}

\begin{figure}[!h]
	\centering
	\includegraphics[width=1.0\columnwidth]{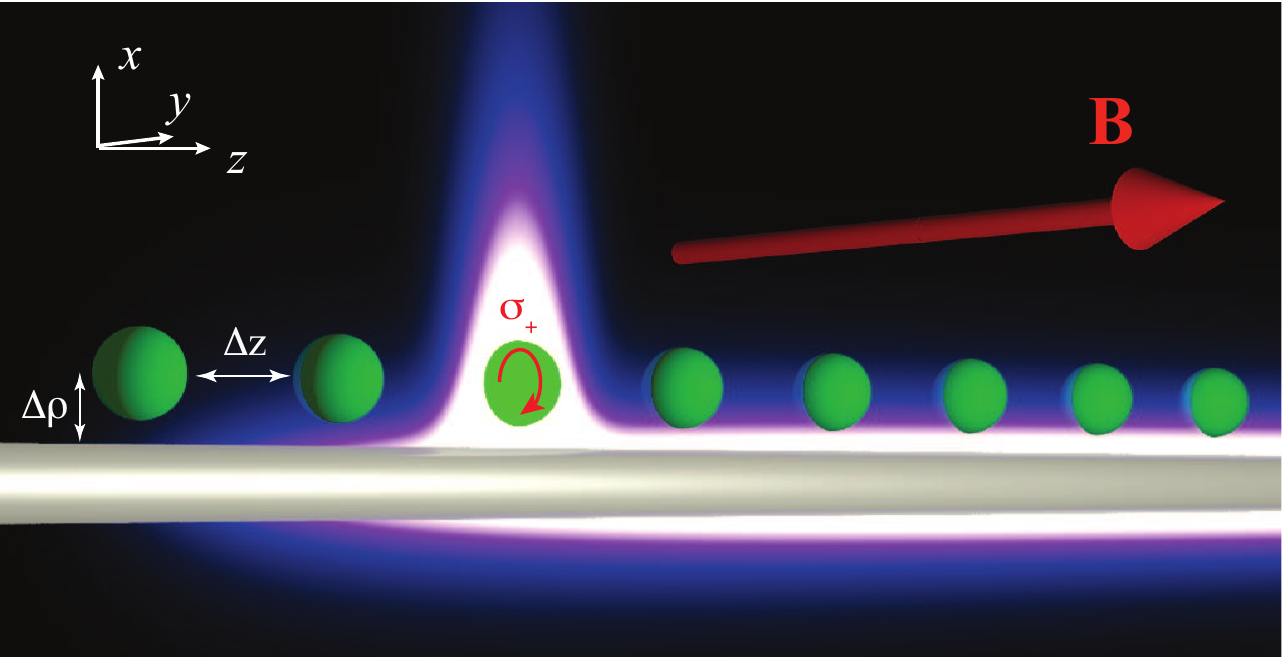}
	\caption{Schematic image of the periodic chain of two level systems on top of a waveguide. Initially, one two level system is excited.The excitation can be transferred either by symmetric short range dipole-dipole or radiative coupling, or by asymmetric long-range coupling via the waveguide mode. {The transverse magnetic field $\mathbf{B}$ breaks the symmetry of coupling of the two-level system to left- and right-propagating waveguide mode.}}
	\label{fig_1}
\end{figure}

The recently emerged field of {\it chiral quantum optics}~\cite{Lodahl2017} promises new possibilities for manipulation and control of quantum states of matter. The  chiral coupling of quantum sources with photonic excitations can be implemented, for example, through the interaction with a topological edge states \cite{Altshuler2013,Ringel2014}. However, one of the most simplest routes for chiral coupling is employment of transverse spin angular momentum of light (SAM), which has recently attracted significant research interest~\cite{Bliokh2015,Bliokh2015a}. In the simplest set-up of an electromagnetic surface or a waveguide mode, the non-zero optical SAM density emerges due to the $\pi/2$ phase shift between the electric field projections onto the interface plane and to its normal~\cite{Bliokh2015}. Important feature of the electromagnetic waves carrying transverse SAM is the spin-momentum locking: the spin projection is defined by the propagation direction of the wave \cite{Rodriguez-Fortuno2013,VanMechelen2016}. This effect, which can be regarded as spin-orbit coupling, has been studied both theoretically and experimentally in many applications related to nano-optomechanics~\cite{Rodriguez-Fortuno2015,Hayat2015,Petrov2016}, topological photonics with surface waves~\cite{Yuen-Zhou2016}, electromagnetic routing~\cite{Kapitanova2014}, and electromagnetically assisted unidirectional spin transfer~\cite{Coles2016} and others. Moreover, the spin-orbit coupling in quasi-one dimensional photonic structures  can be used to engineer new class of quantum information networks~\cite{Pichler2015,Ramos2016}. The basic model under consideration  is an one-dimensional array of two-level systems (TLS) coupled to a quasi-one dimensional photonic nanostructure (See Fig.~\ref{fig_1}).  The current technology allows  for measuring light scattering on  such one-dimensional TLS arrays consisting of thousands cooled atoms trapped near a optical nanofiber \cite{Chang2014}.  Moreover, in Ref.~\onlinecite{PhysRevLett.117.133603,Sorensen2016} it was shown that the  chiral coupling of atom with nanofiber  mode leads to strong modification of   Bragg reflection spectrum. 

In this perspective  the inherent spin-orbit coupling of light in conjunction with the chiral light-matter coupling (which can be achieved be e.g. transverse magnetic field) can allow deterministic transfer of the initial quantum state of the TLS unidirectionally along the channel. Such approach  allows for the engineering of the large scale cascaded quantum networks~\cite{Carmichael1993}, which are immensely in demand in quantum information processing. Despite the importance of this field, the dynamical picture of the excitation transport in an unidirectionally coupled system has not been studied before. 


In this work we focus on the spatio-temporal dynamics of the excited state in such a chiral chain. We adopt the formalism of the Green's function which was proven to be a powerful tool for the studies of quantum dynamics in open systems~\cite{PhysRevLett.117.123901}. We reveal, that under certain approximation the problem of finding the excited states probability amplitude dynamics allows  elegant yet simple analytical solution which agrees well with the rigorous numerical calculations.

\section{Single mode coupling.}
We begin by considering an ensemble consisting of $N$ two-level systems (TLS) forming a one-dimensional linear chain placed parallel to a surface of a photonic/plasmonic nanostructure supporting a single fundamental guided mode. Assuming that coupling is mediated by the guided mode only in the strong spin-locking regime, we formulate the equaitons descibing the dynamics  of the system~\cite{WelschDDC}: 
\begin{equation}
\label{DynEq}
\dot C_n(t)=-i\Omega C_n(t)+\sum_{m=1}^{n-1}G_{nm}C_n(t),
\end{equation}
where $C_n(t)$ is the complex probability amplitude of the $n$-th TLS to be excited at time $t$, the diagonal parameter $ \Omega = \Delta_L + i\gamma_{tot}/2$ contians $\Delta_L$ which  is the Lamb Shift and $\gamma_{tot}$ which is the total spontaneous emission rate consisting of two contributions: emission into radiation and guided modes ($\gamma_{tot} = \gamma_{r} + \gamma_{g}$). The  single mode coupling coefficients $G_{nm}$ between the TLSs with number $m$ and $n$ can be written as $G_{nm}=-\dfrac{\gamma_g}{2}e^{i\phi_{nm}}$,  where $\gamma_g/2$ is the coupling strength, $\phi_{nm}=k^{g}(z_{n}-z_{m})$ is the phase acquired by the photon due to the propagation from emitter $m$ to emitter $n$, and $k^{g}$ is the corresponding propagation constant of the guided mode. We assume strong spin-locking regime, which leads to a unidirectional coupling, i.e. $G_{nm}\neq0$ only for $n>m$. The system of equations ~\eqref{DynEq} can be formulated in the matrix form ${\bf \dot C} (t)={\bf \hat M  C}(t)$, with ${\bf \hat{M}}$ being a lower triangular matrix, which means that the problem is already diagonalized and moreover, it is degenerate. All quantum oscillators have equal transition frequencies and lifetimes and, therefore, the system has only one eigenstate in which the last atom is excited. For this state the corresponding eigenfrequency is complex and a single excitation is not transferred between the atoms, it can only decay to the field modes due to the spontaneous emission process, which is significantly different from the case of symmetric coupling \cite{Kornovan2016}. 

We focus on the problem of the excitation transport through the TLS chain, and for that we consider the initial condition in which the first atom is excited, while all the rest are in the ground state:  $C_1(0)=1,\ C_n(0)=0, n\ge2$. 
Exploiting the triangular form of the matrix $\hat {\bf M}$ and the given form of the initial condition, one can build an exact solution of the problem, which in its compact form can be written as (see Supplementary for the details):  
\begin{align}
C_{n}^{1}(t) = e^{-i\Omega t + i \phi_{n1}}L_{n-1}^{(-1)}(\gamma_g t/2)
\label{laguerre}
\end{align}
\begin{figure}[t]
	\centering
	\includegraphics[width=1.0\columnwidth]{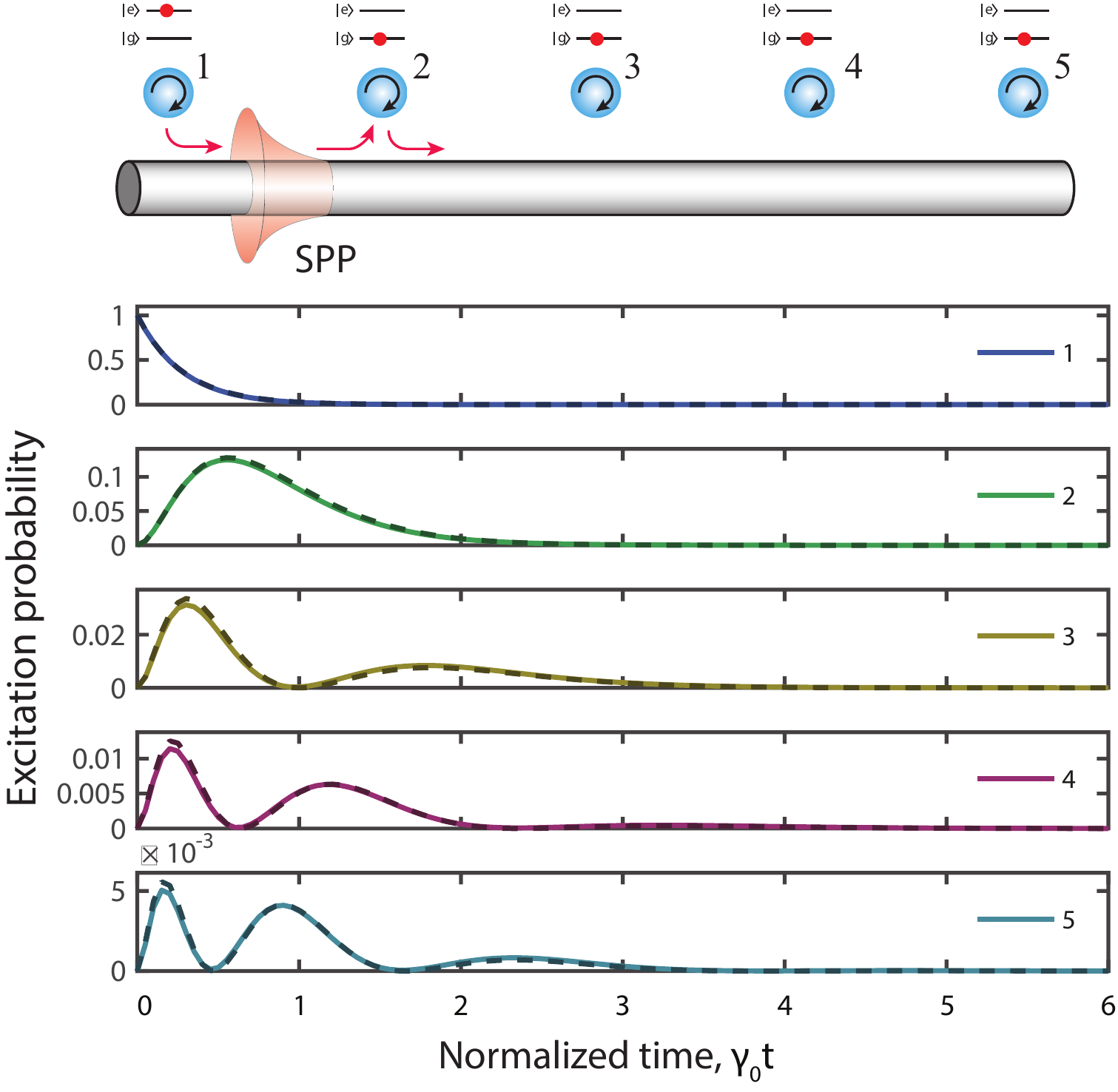}
	\caption{The probabilities for different emitters to be excited at particular time moments $P_n(t)$ for a chain of $N=5$ emitters. The solid and dashed lines are for the numerical and analytical results, correspondingly. For numerical case the probabilities were averaged over $20$ distributions of emitters around their regular positions and the distribution is uniform. The regular separation $\Delta z = 2.0\lambda_{0}$, where $\lambda_0$ is the resonant wavelength of the transition, and the maximal deviation from regular positions $a=\lambda_{pl}/2$. The parameters for the numerical case: nanowire radius is $\rho_c=0.05\lambda_0$, $\varepsilon \approx -16.00 + 0.44i$ and $\Delta\rho = \rho_c$.}
	\label{2DDyn}
\end{figure}
here $L_{n}^{(\alpha)}(x)$ is the generalized Laguerre polynomials of degree $n, \alpha$. This simple solution gives all the insights on the one-directional transport in quantum chains, which we would like to briefly discuss here. First, as it was previously shown~\cite{Ringel2014}, in the case of lossless guided mode the excitation dynamics is {\it irrelevant} of the spatial distribution of the emitters along the $z$-coordinate as $|e^{i\phi_{mn}}|=1$. This makes this system to be tolerant with respect to positional fluctuations, which is a consequence of the perfect one-way transport: the phase of the excitation transported between two emitters $k$ and $l$ always sums up giving the total phase  $k^{g}(z_n-z_1)$. However, this irrelevance of the distribution of the quantum emitters on the final result is also due to the enforced initial condition that only a single (the first) emitter is excited. If we impose a very general initial condition ${\bf C}(t=0)=\left( c_1; c_2; c_3; ...; c_N \right)$, which corresponds to the case when a single excitation is distributed among different atoms meaning that $\sum_{i=1}^{N}|c_i|^2=1$, the answer will depend on the atomic positions. Secondly, the time evolution of the $n$-th atom excitation probability $P_n(t)=|C_{n}(t)|^2$ has trivial exponentially decaying factor $e^{-\gamma t}$, and the stationary solution in such system is 0. Finally, the nontrivial temporal dynamics of the $n$-th emiters' excitation depends on the ampitude of the coupling constant $\gamma_g$ through the corresponding Laguerre's polynamial. 
According to the Laguerre's polynomials  properties \cite{Arfken1972} the number of local excitation maxima for a particular emitter $n$  equals to the number of emitters positioned before it. This dynamics is  shown in Fig.~\ref{2DDyn} (solid lines) for a chain consisting of  $N=5$ emitters.

\section{Metallic Nanowire}
The analytical model we have proposed bases on the interaction of TLSs via arbitrary guided mode. To support these results we consider the interaction of dipole emmiters trhough the plasmonic modes of nanowire. We adopt exact solution of this problem in terms of Green's function approach.  As  we are interested in the probabilities of excitation being transferred from the first emitter of a chain to the $n$-th $P_{n1} = |\langle e_n | \hat U(t,0) | e_1 \rangle|^2$. Here $\hat U(t,0)$ is the evolution operator for our system and $|e_n\rangle$ are the states, where only $n$-th  is excited initially, while all the rest are in the ground state. We can rewrite the matrix elements of the evolution operator according to ~\cite{Cohen-Tannoudji2007}:

\begin{align}
&\langle e_n| \hat U(t,0) |e_1\rangle = \int\limits_C \dfrac{dp}{2\pi i} e^{-ipt/\hbar} \langle e_n| \hat G (p) | e_1 \rangle,
\label{evolop}
\end{align}
where $\hat G (p) = (p - \hat H)^{-1}$ is the resolvent operator of the Hamiltonian $\hat H$ and the contour $C$ here is traversed in the anticlockwise direction and encloses all complex poles of the resolvent (since we consider a subspace containing only discrete states).

We employ the Green's function approach proposed in Ref.~\onlinecite{Gruner1996} in order to quantize the radiation field in the case of absorptive and dispersive media. The electromagnetic field operator in this case reads as $\hat{\mathbf{E}}^{+}(\mathbf{r}) = i \; \sqrt[]{4 \hbar} \int d\mathbf{r^{\prime}} \int\limits_{0}^{\infty} d\omega^{\prime}\frac{{\omega^{\prime}}^2}{c^2} \sqrt{\varepsilon_I (\mathbf{r^{\prime}}, \omega^{\prime})}\mathbf{G}(\mathbf{r}, \mathbf{r^{\prime}}, \omega^{\prime}) \hat{\mathbf{f}}(\mathbf{r^{\prime}}, \omega^{\prime})$ where the bosonic field operators obey the commutation relation $\left[\hat{f}_i(\mathbf{r^{\prime}}, \omega^{\prime}), \hat{f}^{\dagger}_k(\mathbf{r}, \omega)\right] = \delta_{ik}\delta(\mathbf{r^{\prime}} - \mathbf{r})\delta(\omega^{\prime} - \omega)$.

\begin{figure}[t]
	\centering
	\includegraphics[width=1.0\columnwidth]{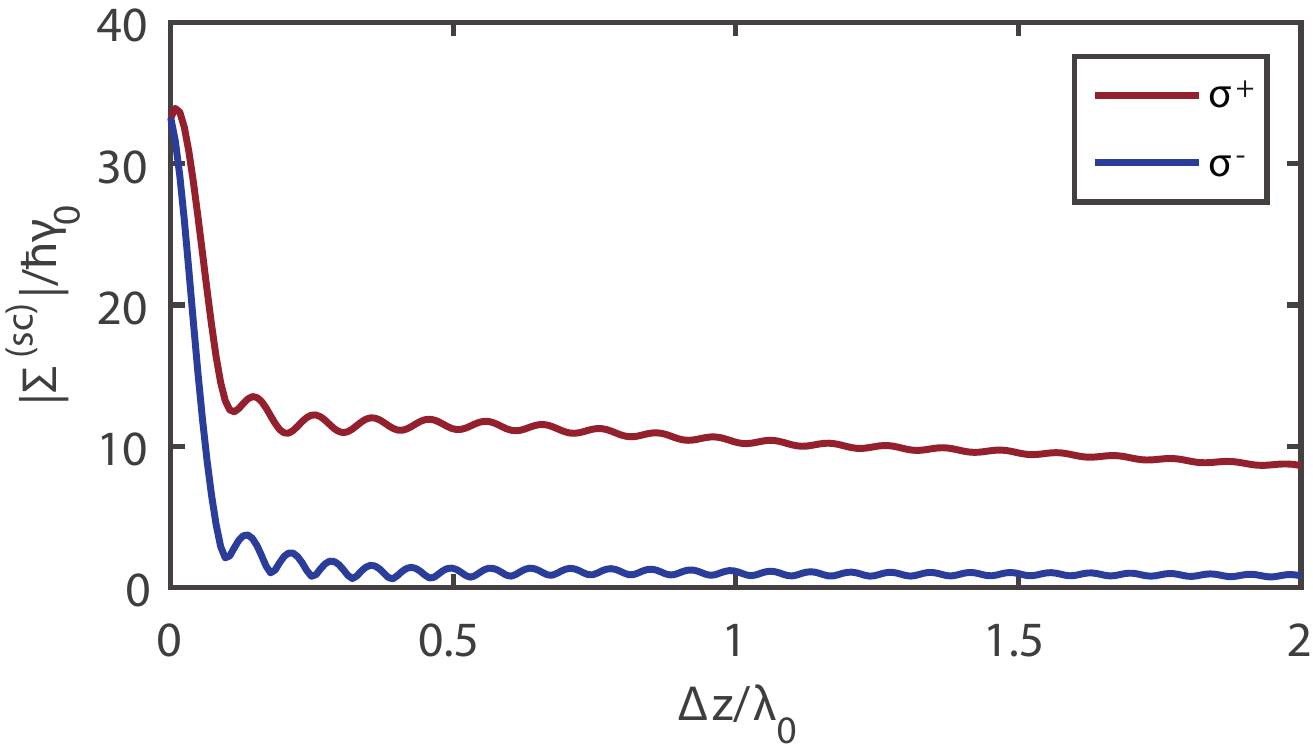}
	\caption{Absolute values of the coupling strength $\Sigma^{(sc)}_{kl}(\hbar \omega_0)$ through the nanowire modes for two atoms spaced by $\Delta z$ and placed at a distance $\Delta\rho$ from the surface of the metallic nanowire measured in $\hbar \gamma_0$, where $\gamma_0$ is a free-space spontaneous emission rate. The quantization axis is chosen to be $\mathbf{e_y}$, therefore, $\mathbf{e_{\sigma_+}}=-(i\mathbf{e_{x}+\mathbf{e_{z}}})/\sqrt{2}$, $\mathbf{e_{\sigma_-}}^{ }=-\mathbf{e_{\sigma_+}^{*}}$. The parameters $\rho_c$, $\Delta\rho$, and $\varepsilon$ are the same as for Fig. \ref{2DDyn}.}
	\label{2aDDInt}
\end{figure}

%

We then find the projections of  the resolvent operator on states with a single atomic excitation which are given by~\cite{Cohen-Tannoudji2007}:

\begin{align}
&\hat P \hat G(p) \hat P = \hat P \dfrac{1}{p - \hat H_0 - \hat \Sigma(p)}\hat P, \label{Projection1}\\
&\hat \Sigma(p)=\hat V + \hat V \hat G(p) \hat V \approx \hat V + \hat V \hat G_0(p=\hbar\omega_0) \hat V,\label{ProjectionApprox}
\end{align}
where $\hat P = \sum_{j=1}^{N}|e_{j}\rangle \langle e_{j}|$ is the projection operator onto the corresponding subspace, $\hat H_0$ is the unperturbed Hamiltonian and $\hat \Sigma(p)$ is the level-shift operator~\cite{Cohen-Tannoudji2007}, known also as self-energy part, which provides us with the correction to the unperturbed Hamiltonian $\hat H_0$ due to the interaction between the quantum emitters.
The exact form of the projected resolvent operator is shown in Eq.~\eqref{Projection1}, while Eq.~\eqref{ProjectionApprox} imposes two approximations: i) we limit ourselves in the calculation of the self-energy only up to the second order in $\hat V$;  ii) we consider a near-resonant interaction between TLS computing $\hat G$ at the resonant frequency $p=\hbar \omega_0$. Within these approximations the matrix elements $\langle e_k|\hat \Sigma (p)|e_l\rangle = \Sigma_{kl}(p)$ show the coupling strength of two emitters with numbers $k$ and $l$. It is defined by the electromagnetic Green's function of the system \cite{Dung2002} and reads as: $\Sigma_{kl}(\omega_0)=-4\pi k_0^2\mathbf{d_{k}^{*}}\mathbf{G}(\mathbf{r_k}, \mathbf{r_l},\omega_0)\mathbf{d_l}$, where $k_0=\omega_0/c$, $\mathbf{d_k}$ is the transition dipole moment and $\mathbf{G}(\mathbf{r_k}, \mathbf{r_l},\omega_0)$ is the classical Green's tensor. By taking the exact Green's function of a metallic nanowire (see Supplementary), we have studied
the interaction strength between two emitters mediated by the propagating surface plasmon-polariton modes (SPP). By introducing an external magnetic field along $y$-axis one can achieve efficient coupling of the emitters with circular transitions only with dipole moment ${\bf d}_{\sigma_+}=-d_0(i{\bf e}_x+{\bf e}_z)/\sqrt{2}$. Due to the spin-locking of the $\sigma_+$ transition with SPP mode the coupling strength between the emmiters is strongly asymetric as one can see from the Fig.~\ref{2aDDInt}. %
The considered nanowire has $\varepsilon \approx -16 + 0.44i$, which corresponds to the silver permittivity at $\lambda_0 = 600$ nm \cite{JC1972}, the nanowire radius $\rho_c = 0.05\lambda_0\approx 30$ nm, and the distance from the fiber surface is $\Delta\rho = \rho_c$.
As the distance between the TLS increases, the interaction through the guided mode of the wire play the dominant role. Any visible oscillations occur due to the interference between the fundamental guided mode and higher-order radiation modes. As can be seen clearly, for such a thin fiber, supporting only one fundamental guided mode with radial eigenvalue $n=0$, the interaction strength is very different for the transition dipole moments rotating in the opposite directions ($\sigma_+$ and $\sigma_-$). Though the TLS transition frequency is far from the SPP resonance the asymmetry of the coupling strength is on the order of 10. This allows us to apply the unidirectional model and compare this to the solution of the numerical one, which is shown in Fig.~\ref{2DDyn} (dashed lines). We get almost perfect correspondence between our simplified model (solid line) and the obtained numerical solution (dashed line), which confirms that all the unique properties of the unidirectional transport formulated before can be observed in realistic structures.


\begin{figure}[t]
	\centering
	\includegraphics[width=1.0\columnwidth]{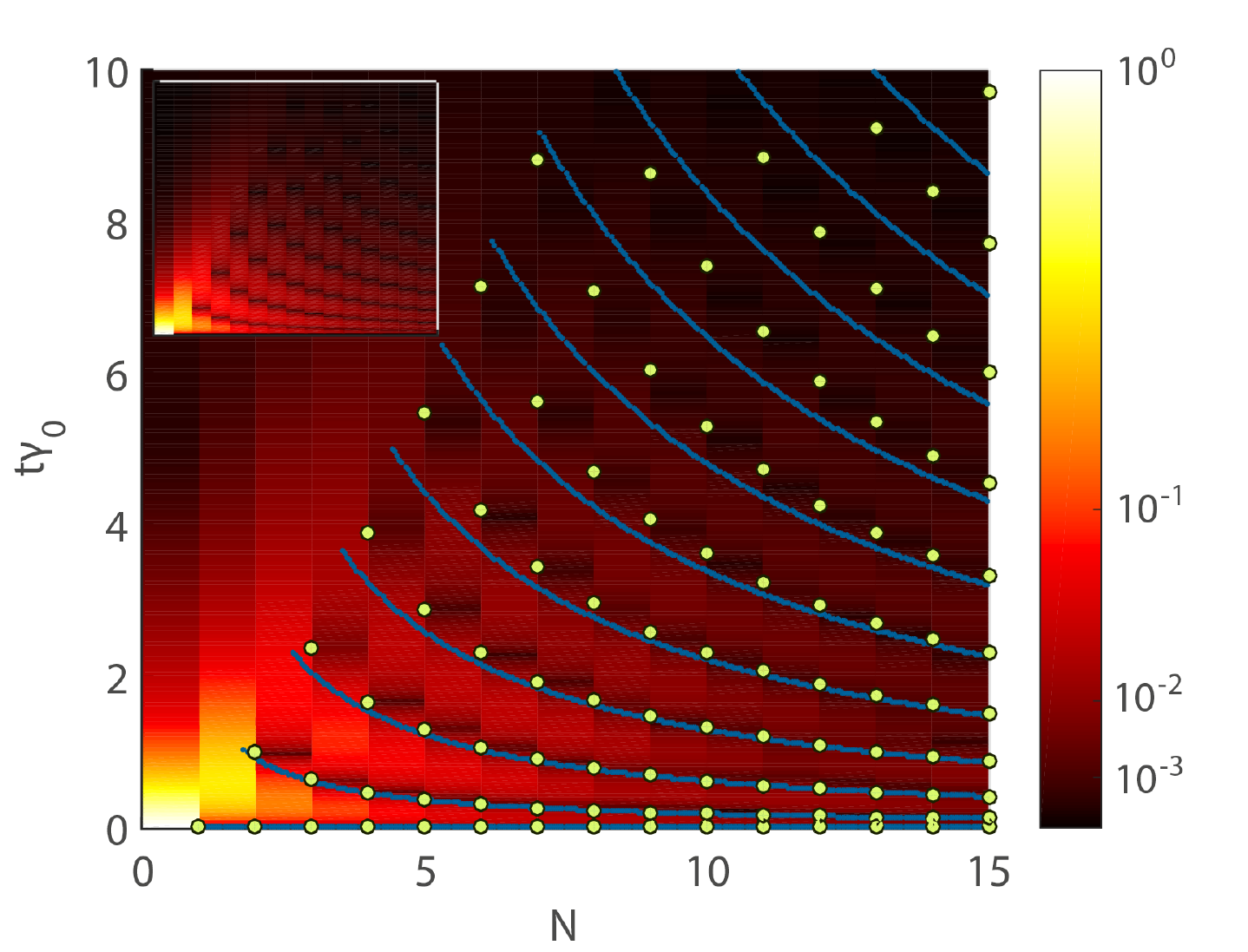}
	\caption{Distribution of excitation between different emitter $n$ in a chain of total $N=15$ emitters. In a computational model parameters $\Delta z$, $\Delta\rho$, $\rho_c$, and $\varepsilon$ are the same as for Fig. \ref{2DDyn}. Yellow circles correspond to exact positions of zeros in dynamics for each emitter for the case of a perfect unidirectionality, while the dark blue lines are zeros of $J_{1}(2\sqrt{(N-1)\gamma_gt/2})$. The inset figure shows analytical results calculated with Eq. \eqref{laguerre}.}
	\label{3DDynamics}
\end{figure}


Now if we plot the excitation probability versus both time $t$ and emitter number $N$ it can observed that there are excitation waves propagating in the chain, which can be indicated by tracking the positions of zeros in dynamics; the fronts of these waves can be described with the help of asymptotic relation connecting the Laguerre polynomials and the Bessel functions for large $n$ and fixed time $t$: $L_{N}^{(\alpha)}(\gamma_gt/2) \approx N^{\alpha/2} e^{\frac{\gamma_gt}{4}} \frac{J_{\alpha}(2\sqrt{N\gamma_gt/2})}{(\gamma_gt/2)^{\alpha/2}}$ and zeros of this function were plotted for the case of continuous $N$ in Fig. \ref{3DDynamics}. Notice that since zeros of the Bessel function are constants, these waves propagate with negative phase velocity as for a larger emitter number $N$ zeros appear at earlier times $t$.


\section{  {Collective Emission}} 
In the previous chapters we have considered the redistribuiton of a single excitation initially localized at the first atom in the chain. However, the physical mechanisms lying beyond the emission of specially prepared states could be of interest as well. One of the common cases is the Dicke state having superradiant property. The collective emission of excitation in the case of unidirectional coupling can significantly differ from symmetrical coupling. 

\begin{figure}[t]
	\centering
	\includegraphics[width=1.0\columnwidth]{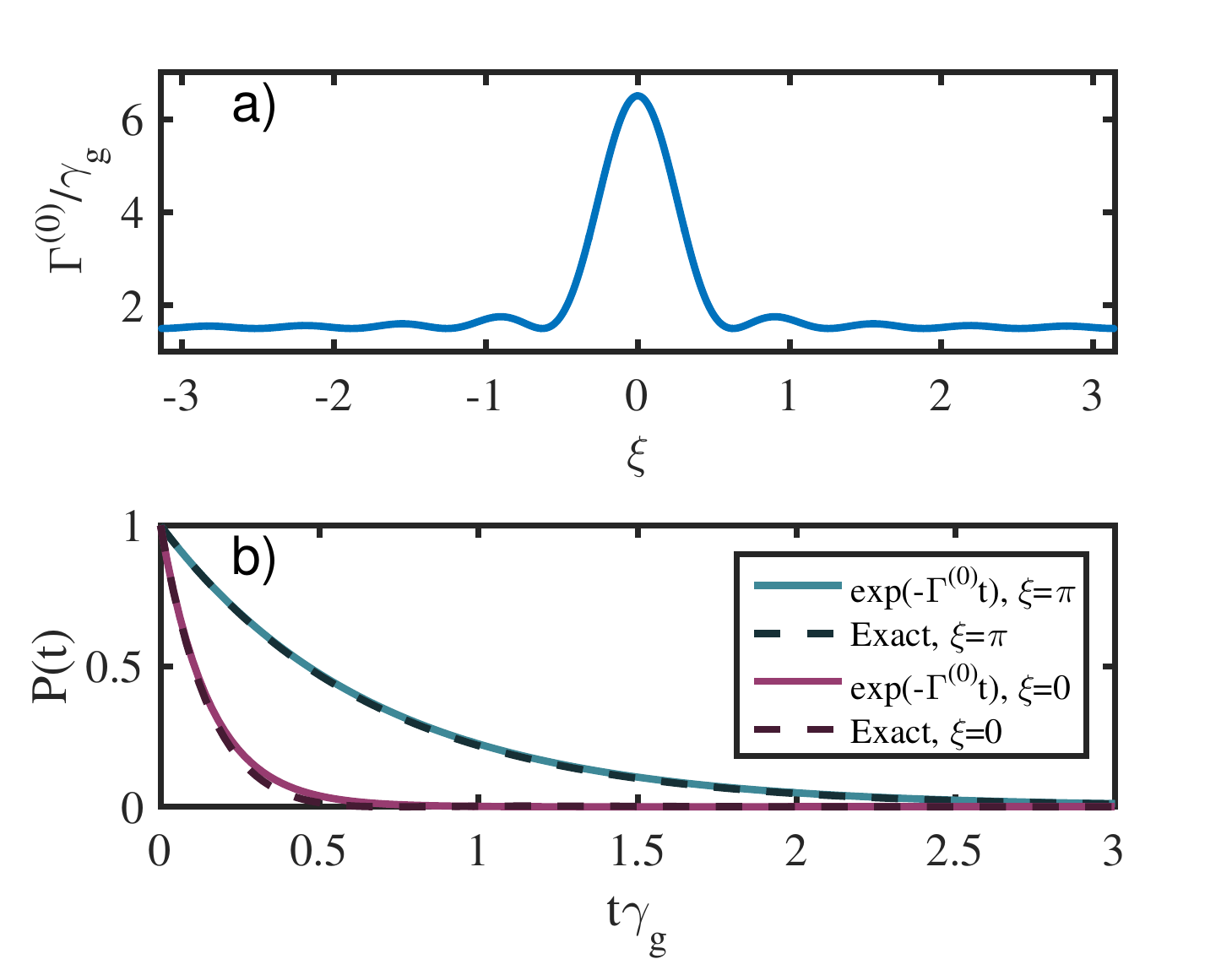}
	\caption{a) Dependence of the initial spontaneous emission rate $\Gamma^{(0)}$ on $\xi$ for a chain of $N=10$ atoms, when $\gamma_g = \gamma_r = 1.0$. b) Dynamics for the case of out of phase  (blue) and in phase (purple) neighboring emitters. Solid lines represent the exact solution, dashed - exponential with $\Gamma^{(0)}$ given by \eqref{gammainit}. The parameters are the same as for a).}
	\label{SR}
\end{figure}

For such a collective state  in the absence of retardation the dynamics can be found as 
\begin{eqnarray}
	C(t) = \mathbf{C_{init}}^\dagger\mathbf{U(t)}\mathbf{C_{init}},
\end{eqnarray}
with $U_{k,l}(t) = e^{-i\Omega t}L^{(-1)}_{k-l}(\gamma_gt/2) e^{i\phi_{k,l}}$ being a lower triangular matrix (similar to \eqref{evolop}) of probability amplitudes for $k$-th atom to be excited at time $t$, while initially only the $l$-th atom was excited, and $C_{init,l} = \dfrac{e^{i\psi_l}}{\sqrt{N}}$ is a vector of the initial condition. We proceed by considering that atoms in our chain are spaced regularly $\phi_{k,l}=(k-l)\phi$ and that $\psi_l = (l-1)\psi$, where both $\phi$ and $\psi$ are purely real. General answer for the complex phase differences is presented in  the Supplementary. In this case it can be found that $C(t) = \dfrac{e^{-i\Omega t}}{N}\sum\limits_{k=1}^{N}(N-(k-1))e^{i(k-1)\xi}L^{(-1)}_{k-1}(\gamma_gt/2)$ with $\xi = \phi - \psi$. Next we consider sufficiently small times and expand $C(t)$ to the first order in $t$ finally obtaining

\begin{eqnarray}
C(t\to 0)\sim  1 - \left[i\Omega  + \dfrac{\gamma_g}{2} \dfrac{e^{i\xi}\left( N+e^{iN\xi}-Ne^{i\xi}-1\right)}{N\left(e^{i\xi} - 1\right)^2} \right]t \sim\nonumber\\
 1 - \dfrac{\Gamma^{(0)}}{2}t,\qquad
 \label{gammainit}
\end{eqnarray}
where $\Gamma^{(0)}$ is the initial spontaneous emission rate being a real part of the expression in square brackets. Its dependence upon $\xi$ is illustrated in Fig. \ref{SR}, a).

It is reasonable to proceed by considering the two cases corresponding to the situations when the neighbouring atoms are emitting photons in- and out of phase:

\begin{eqnarray}
&\dfrac{\Gamma^{(0)}}{2} =  \begin{cases}
-i\Omega - \dfrac{\gamma_g}{2}\dfrac{(N - 1)}{2}, &\text{if $\xi = 2\pi m$;} \\
-i\Omega + \dfrac{\gamma_g}{2}\dfrac{(2N - 1 + e^{iN\pi})}{4N}, & \text{if $\xi = \pi (2m + 1)$}.
\end{cases}
\end{eqnarray}

Notice that for even $N$ in the second case the coefficient $\Gamma^{(0)}$ reaches its absolute minimum. In the limit of strong coupling with the guided mode $\gamma_g \gg \gamma_r$ and large emitter number $N \gg 1$ for the $\xi = 2\pi m$ case $\Gamma^{(0)} = \dfrac{N\gamma_g}{2}$ unlike the  $N\gamma_g$ factor known for the emission of the symmetric Dicke superradiant state \cite{Dicke1954}. For the out-of phase case when $\xi = 2\pi(m+1)$ the initial decay rate is $\Gamma^{(0)} =  \dfrac{\gamma_g}{2}$. The dynamics for both situations are illustrated in Fig. \ref{SR}, b).

Concluding, we have proposed a simple analytical model of the unidirectional quantum transport mediated by spin-locked coupling to an arbitrary waveguide mode. We have obtained the exact analytical solution, showing that the dynamics of the TLS is described by the Laguerre polynomials. The behaviour of the chiral TLS system is fully defined by the amplitude of the coupling coefficient of a single emitter with the waveguide mode. From the obtained solution it immediately follows, that unidirectional system possesses the tolerance with respect to the positional disorder. 
Our model also predicts that for systems with perfectly asymmetric coupling, the symmetric Dicke superradiant state in a special case of phase-matched positions of the emitters has the emission rate equal to $N\gamma/2$, contrary to a value $N\gamma$ typical for systems with symmetrical interaction. In order to verify our model, we have performed the simulation of quantum excitation transfer through SPP mode of a metallic nanowire, constructing the evolution operator, basing on the exact electromagnetic Green's function of the system. The exact simulations have shown good agreement with  the proposed analytical model, which allows its application for describing any chiral quantum system unidirectionally coupled to a waveguide mode. 

\section*{Acknowledgement}
We are grateful to Mikhail Glazov, Andrey Bogdanov, and  Yuri Kivshar for helpful discussions. The work was supported by the Russian Foundation for Basic Research proj. \# 16-32-60167. 

\bibliographystyle{apsrev4-1}
\bibliography{unibibio}
\end{document}